\documentclass[aps,pra,twocolumn,superscriptaddress,nobalancelastpage,amsmath,amssymb]{revtex4-2}
\usepackage{graphicx}
\graphicspath{{/}{figures/}}
\usepackage{bm}
\usepackage{hyperref}

\newcommand {\ii}      {\mathrm{i}}

\newcommand {\nm}      {\,\mathrm{nm}}
\newcommand {\meV}     {\,\mathrm{meV}}
\newcommand {\abs}[1]  {\lvert#1\rvert}

\renewcommand {\vec}   {\mathbf}

\newcommand {\ket}[1]      {\lvert#1\rangle}

\newcommand {\bramidket}[3] {\langle#1\vert#2\vert#3\rangle}

\newcommand {\avg}[1]   {\langle#1\rangle}

\newcommand {\half}       {\tfrac{1}{2}}
\newcommand {\threehalf}  {\tfrac{3}{2}}
\newcommand {\spinup}  {\mathord\uparrow}
\newcommand {\spindn}  {\mathord\downarrow}
\newcommand {\dpg}[2]  {\tilde{#1}_{#2}}

\newcommand {\isopz}   {\tilde{\mathcal{P}}_z}
\newcommand {\eqn}     {Eq.~}
\newcommand {\kdotp}   {\ensuremath{\vec{k}\cdot\vec{p}}}

\newcommand {\idmat}   {\mathbf{1}}
\newcommand {\tildevecj} {\tilde{\textbf{\j}}}
\newcommand {\Hstr}    {H^{\mathrm{str}}}

\DeclareMathOperator{\sgn}{sgn}
\DeclareMathOperator{\diag}{diag}
\DeclareMathOperator{\tr}{tr}

\begin{document}

\date{\today}

\title{Parity symmetry as the origin of `spin' in the quantum spin Hall effect}
\author{Wouter Beugeling}
\affiliation{Experimentelle Physik III, Physikalisches Institut, Universit\"at W\"urzburg, Am Hubland, 97074 W\"urzburg, Germany}
\affiliation{Institute for Topological Insulators, Am Hubland, 97074 W\"{u}rzburg, Germany}

\begin{abstract}
  The quantum spin Hall effect arises due to band inversion in topological insulators, and has the defining characteristic that it hosts helical edge channels at zero magnetic field, leading to a finite spin Hall conductivity. The spin Hall conductivity is understood as the difference of the contributions of two spin states.
  In the effective four-band BHZ model, these two spin states appear as two uncoupled blocks in the Hamiltonian matrix.
  However, this idea breaks down if additional degrees of freedom are considered. The two blocks cannot be identified by proper spin $S_z$ or total angular momentum $J_z$, both not conserved quantum numbers. 
  In this work, we discuss a notion of block structure for the more general $\kdotp$ model, defined by a conserved quantum number that we call \emph{isoparity}, a combination of parity $z\to-z$ and spin.
  Isoparity remains a conserved quantity under a wide range of conditions, in particular in presence of a perpendicular external magnetic field. From point-group considerations, isoparity is fundamentally defined as the action of $z\to-z$ on the spatial and spinorial degrees of freedom.
  Since time reversal anticommutes with isoparity, the two blocks act as Kramers partners. The combination of conductivity and isoparity defines spin conductivity.
  This generalized notion of spin Hall conductivity uncovers the meaning of `spin': It is not the proper spin $S_z$, but a crystal symmetry that is realized by a spinorial representation.
\end{abstract}

\maketitle

\section{Introduction}
\label{sec_intro}

The experimental realization \cite{KonigEA2007,BruneEA2011} and theoretical prediction \cite{KaneMele2005PRL95-14,BernevigEA2006} of the quantum spin Hall effect have initiated a huge interest in the physics of topological insulators in the past two decades.
As opposed to the quantum Hall effect, that is generated by an external magnetic field \cite{ThoulessEA1982}, the quantum spin Hall effect is rooted in the intrinsic band inversion of the material \cite{BernevigEA2006,KonigEA2007}. The quantum spin Hall effect is tied to time-reversal symmetry: In the absence of an external magnetic field, the Hall conductance vanishes, but at the same time the system hosts a pair of counterpropagating edge channels, giving rise to so-called spin Hall conductance \cite{BernevigEA2006}. In practice, these counterpropagating edge channels can be observed as a quantized longitudinal conductance $G=2e^2/h$ in the absence of a magnetic field \cite{KonigEA2007,RothEA2009}.

The mechanism of band inversion giving rise to helical edge channels has been elegantly explained in the model proposed by Bernevig, Hughes, and Zhang (BHZ) \cite{BernevigEA2006}.
In order to explain the transition from trivial to topological behaviour in HgTe/(Hg,Cd)Te quantum wells at the critical thickness $d_c=6.3\nm$, it is sufficient to project the Hamiltonian to the subspace of the four subbands $\ket{\mathrm{E}1,\pm}$ and $\ket{\mathrm{H}1,\pm}$, the spin-degenerate pairs of the first subband in both the conduction and valence band. The two spin states are uncoupled, so that the $4\times4$ matrix Hamiltonian splits into two $2\times 2$ blocks \cite{BernevigEA2006}.

The BHZ model is obtained by an expansion of the Hamiltonian near the $\Gamma$ point projected to the four subband states of its basis. It is extracted from $\kdotp$ models which contain many more degrees of freedom and describe more aspects of the band structure in an accurate manner \cite{Kane1957,Weiler1981bookchapter,PfeufferJeschke2000_thesis,NovikEA2005,Winkler2003_book}. The four-band BHZ model is unable to capture physics where these other degrees of freedom are essential.
For example, it does not predict the dispersion of the \emph{camel back}, a maximum in the dispersion of the valence band at finite momentum, that arises due to hybridization of the $\mathrm{E}1$ subband with higher order heavy-hole subbands like $\mathrm{H}2$ \cite{OrtnerEA2002,ShamimEA2020SciAdv}.
The camel back can play a dominant role in the transport properties of HgTe quantum wells of thickness $d\gtrsim 7\nm$, already quite close to $d_c$. A proper description thus requires band structures obtained from the $\kdotp$ model \cite{ShamimEA2020SciAdv} or from extended BHZ models with more than four subbands \cite{KrishtopenkoTeppe2018PRB}.

The necessity of considering these more advanced models motivates the question whether they admit a similar block structure as the four-band BHZ model. Extended BHZ models do, but the grouping of the subbands is not entirely intuitive: The subband $\ket{\mathrm{H}2,+}$ couples with $\ket{\mathrm{E}1,-}$ and $\ket{\mathrm{H}1,-}$ \cite{KrishtopenkoTeppe2018PRB}, where the label $\pm$ refers to $\avg{\sgn(J_z)}$, where $J_z$ is total angular momentum with quantization axis along the growth direction $z$.
This example shows that the blocks are not defined by spin only, but by a combination of spin and parity: The envelope function $g(z)$ of $\ket{\mathrm{H}2,\pm}$ is odd under parity $z\to-z$, whereas that of $\ket{\mathrm{H}1,\pm}$ is even \cite{RotheEA2010}. The idea of a discrete symmetry that acts on spin and parity $z\to-z$ simultaneously extends to the $\kdotp$ model as well \cite{AndreaniEA1987}.

In this work, we elaborate on the role of the combination of parity and spin, called \emph{isoparity} $\isopz$, in defining a block structure in the $\kdotp$ model in the inversion symmetric approximation \cite{[{Bulk inversion asymmetric terms have found to be small, see, e.g., }][{.}]OrlitaEA2011}. Isoparity is a conserved quantity with respect to the $\kdotp$ Hamiltonian $H_k$ (i.e., $[H_k,\isopz]=0$) even at finite momentum, unlike spin $S_z$ and total angular momentum $J_z$. Uniaxial strain, encoded by the Pikus-Bir strain Hamiltonian \cite{BirPikus1974_book,NovikEA2005}, and couplings with the magnetic fields in the perpendicular direction ($\vec{B}\parallel \hat{z}$) also preserve isoparity.
While isoparity is not conserved if the symmetry $z\to -z$ is broken by, e.g., a perpendicular electric field (gate voltage) or in an asymmetric geometry like a type-II topological insulator device \cite{LiuEA2008PRL100}, it can be convenient to define the basis in terms of isoparity eigenstates; the inversion asymmetric terms then appear in the off-diagonal block, analogous to the structure known in the context of the BHZ model \cite{RotheEA2010,LiuEA2008PRL100}.

From analysis of the representations of the relevant double point group (in the inversion symmetric model, $\dpg{O}{h}$ or $\dpg{D}{4h}$ for bulk and quantum well geometry, respectively), we confirm that isoparity is identical to the mirror transformation $m_z$ in the $z$ direction, up to the imaginary factor $\ii$.
The block-like structure of the Hamiltonian, given by invariance under isoparity, is thus fundamentally a consequence of crystal symmetry.
By identifying the action of time reversal in the context of representation theory, we show that isoparity and time reversal anticommute. Time reversal thus maps one block to the other. Importantly, this relation proves that the two blocks are Kramers partners, establishing it as generalization of this idea known from the BHZ model \cite{BernevigEA2006}.
The combination of the ordinary conductance tensor $\sigma_{ij}$ with isoparity yields the spin conductance $\tilde\sigma_{ij}$. By construction, spin conductance is odd under time reversal, unlike ordinary conductance which is even; these properties are dictated by crystal and time reversal symmetry as well as Onsager's relation \cite{Grimmer1993,Grimmer2017}.
This difference is related to the fundamentally different origins of the Hall effect and spin Hall effect, respectively \cite{BottcherEA2020PRB}.

With these ingredients, we establish an unambiguous interpretation of the word `spin' in `quantum spin Hall effect': It does not indicate proper spin $S_z$, which is not a conserved quantum number, but it refers to the crystal symmetry $m_z$. The name `spin' is appropriate nevertheless, as it involves an operation that is inherently spin-like in nature, namely, it acts according to a spin representation onto the half-integer angular momentum degrees of freedom and it anticommutes with time reversal.

The outline of the paper is as follows. In Sec.~\ref{sec_modelling}, we briefly introduce our model, with further details in the Appendix.
In Sec.~\ref{sec_isoparity}, we provide the ideas of the splitting into blocks in more detail, and define isoparity based on those observations. We then prove that isoparity is invariant, and how external effects (magnetic, electric, and strain couplings) affect this invariance.
In Sec.~\ref{sec_pointgroup}, we interpret isoparity in the context of spin representations of the point group.
We discuss time reversal in Sec.~\ref{sec_timereversal}.
The notions of conductivity and spin conductivity are defined and analyzed in Sec.~\ref{sec_cond_isocond}.
The concluding Sec.~\ref{sec_conclusion} discusses some practical applications and implications for the interpretation of the term `spin' in this framework.

\section{Modelling}
\label{sec_modelling}

As the basis for our analysis, we use the $8$-orbital $\kdotp$ model for (Hg,Cd)Te and (Hg,Mn)Te presented in Refs.~\cite{Weiler1981bookchapter,PfeufferJeschke2000_thesis,NovikEA2005} as a starting point. The $8$-orbital basis is
{\allowdisplaybreaks
\begin{align}
  \ket{\Gamma_6,+\half}&=\ket{S}\otimes\ket{\spinup},\nonumber\\
  \ket{\Gamma_6,-\half}&=\ket{S}\otimes\ket{\spindn},\nonumber\\
  \ket{\Gamma_8,+\threehalf}&=\tfrac{1}{\sqrt{2}}\ket{X+iY}\otimes\ket{\spinup},\nonumber\\
  \ket{\Gamma_8,+\half}&=\tfrac{1}{\sqrt{6}}\left(\ket{X+iY}\otimes\ket{\spindn}-2\ket{Z}\otimes\ket{\spinup}\right),\nonumber\\
  \ket{\Gamma_8,-\half}&=\tfrac{1}{\sqrt{6}}\left(-\ket{X-iY}\otimes\ket{\spinup}-2\ket{Z}\otimes\ket{\spindn}\right),\nonumber\\
  \ket{\Gamma_8,-\threehalf}&=-\tfrac{1}{\sqrt{2}}\ket{X-iY}\otimes\ket{\spindn},\nonumber\\
  \ket{\Gamma_7,+\half}&=\tfrac{1}{\sqrt{3}}\left(\ket{X+iY}\otimes\ket{\spindn}+\ket{Z}\otimes\ket{\spinup}\right),\nonumber\\
  \ket{\Gamma_7,-\half}&=\tfrac{1}{\sqrt{3}}\left(\ket{X-iY}\otimes\ket{\spinup}-\ket{Z}\otimes\ket{\spinup}\right),
  \label{eqn_kdotpbasis}
\end{align}%
}%
in terms of the s-orbital labelled $S$ and the p-orbitals labelled $X$, $Y$, and $Z$, as well as the two spin states $\spinup$ and $\spindn$. For convenience of notation, we split the Hamiltonian $H$ into orbital blocks $H_{pq}$ with $p,q=6,7,8$ labelling the orbital sectors $\ket{\Gamma_p, \cdot}$,
\begin{equation}\label{eqn_ham_blocks}
  H = \left(\begin{array}{c|c|c}
        H_{66}&H_{68}&H_{67}\\ \hline
        H_{86}&H_{88}&H_{87}\\ \hline
        H_{76}&H_{78}&H_{77}
      \end{array}\right).
\end{equation}
Hermiticity of $H$ implies that $H_{qp}=H_{pq}^\dagger$. Our investigation will mostly be targeted at the inversion symmetric part of the $\kdotp$ Hamiltonian, denoted $H_k$. The blocks of $H_k$ are given in Appendix~\ref{sec_app_hamkdotp}, together with the matrix representations of spin $S_i$ and total angular momentum $J_i$ ($i=x,y,z$).
Strain, magnetic and electric couplings, as well as inversion asymmetric terms will be discussed separately.

\section{Spin, parity, isoparity}
\label{sec_isoparity}

\subsection{Incentive}
\begin{figure}
  \includegraphics{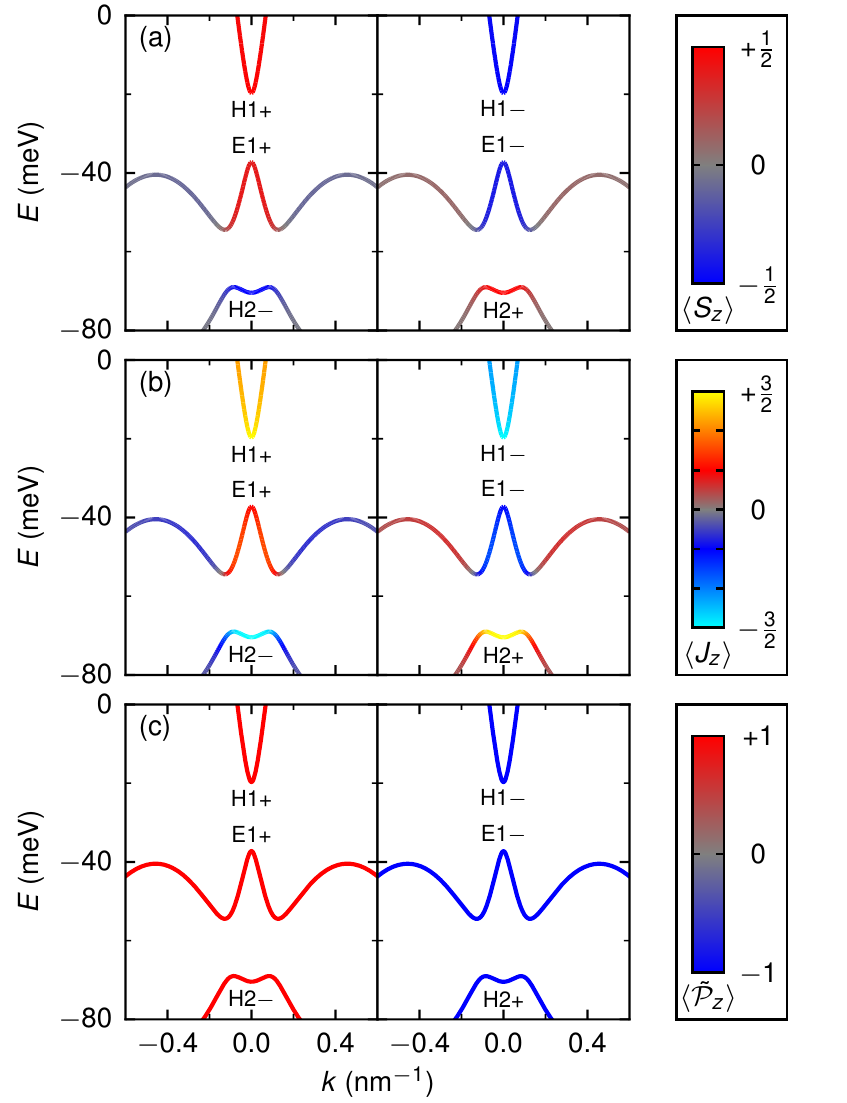}
  \caption{Dispersion for a $7\nm$ HgTe quantum well between Hg$_{1-x}$Cd$_x$Te ($x=0.68$) barriers in the inversion symmetric approximation. The momentum is parametrized as $(k_x,k_y)=k(1/\sqrt{2},1/\sqrt{2})$, along the (110) crystal direction. (a) Colours indicate spin expectation value $\avg{S_z}$. The visible subbands (labelled) belong to the same block; the states of the other block are degenerate, but not shown. (b) Same, with colours indicating $\avg{J_z}$. (c) Colours indicate $\avg{\isopz}$. We show the two degenerate copies of each state separately (left/right column), with the subbands grouped according to a $6$-band BHZ type model (see Introduction and \eqn\eqref{eqn_bhzblocks}).}
  \label{fig_disp_colors}
\end{figure}

The Hamiltonian $H_k$ does not preserve spin $S_z$ or total angular momentum $J_z$,
\begin{equation}
  [H_k,S_z]\not= 0,\qquad [H_k,J_z]\not= 0,
\end{equation}
for $\vec{k}\not=0$. 
This can be demonstrated straightforwardly by examining the eigenstate expectation values $\avg{S_z}$ and $\avg{J_z}$ in a typical quantum well dispersion, e.g., for $7\nm$ thick HgTe between (Hg,Cd)Te barriers, as illustrated in Fig.~\ref{fig_disp_colors}. In Fig.~\ref{fig_disp_colors}(a) and (b), we have plotted the dispersion with the colouring based on the expectation values $\avg{S_z}$ and $\avg{J_z}$, respectively. The curves not being uniformly coloured indicates that these expectation values are momentum-dependent. Thus, $S_z$ and $J_z$ do not define good quantum numbers.


The widespread terminology `spin up' and `spin down' states (with notation typically $+$ and $-$, respectively, sometimes $\spinup$ and $\spindn$, respectively) is typically understood in the sense $J_z > 0$ and $J_z < 0$; the word `spin' should be read as `pseudospin'. Whereas this is an intuitive and workable definition in some limited contexts, in particular in the four-band BHZ model \cite{BernevigEA2006}, it can be problematic in other cases that require consideration of more degrees of freedom. Firstly, for large in-plane momentum $(k_x, k_y)$, hybridization may occur between subbands, e.g., between the subbands labelled $\mathrm{E}1$ and $\mathrm{H}2$ in Fig.~\ref{fig_disp_colors}, leading to the typical camel back feature \cite{ShamimEA2020SciAdv}. The hybridization is strong enough that $\avg{J_z}$ changes sign at $k\approx0.15\nm^{-1}$, as shown by the colour of the band labelled $\mathrm{E}1+$ in Fig.~\ref{fig_disp_colors}(b) changing from red to blue. Secondly, extension of the original $4$-band BHZ model with additional subbands cannot be done in a naive manner: For example, if one adds the second heavy-hole subbands $\ket{\mathrm{H}2,\pm}$ to the `spin up' and `spin down' block of the $4$-band BHZ model, the two `blocks' are grouped \cite{RotheEA2010,KrishtopenkoTeppe2018PRB}
\begin{align}
  &\{\ket{\mathrm{E}1,+}, \ket{\mathrm{H}1,+}, \ket{\mathrm{H}2,-}\},\nonumber\\
  &\{\ket{\mathrm{E}1,-}, \ket{\mathrm{H}1,-}, \ket{\mathrm{H}2,+}\},
  \label{eqn_bhzblocks}
\end{align}
where the $\mathrm{H}2$ bands are in the opposite block from what one could naively expect. Thus, in extended BHZ-like models, one is no longer able to label the two blocks by `spin' meaning $\avg{\sgn(J_z)}$. Likewise, for the $\kdotp$ model, containing many more degrees of freedom, this notion of `spin up' and `spin down' blocks cannot work properly. The notions of spin defined by $S_z$, $J_z$, and $\sgn(J_z)$ not being good quantum numbers makes them unsuitable for labelling the blocks.


\begin{figure}
  \includegraphics{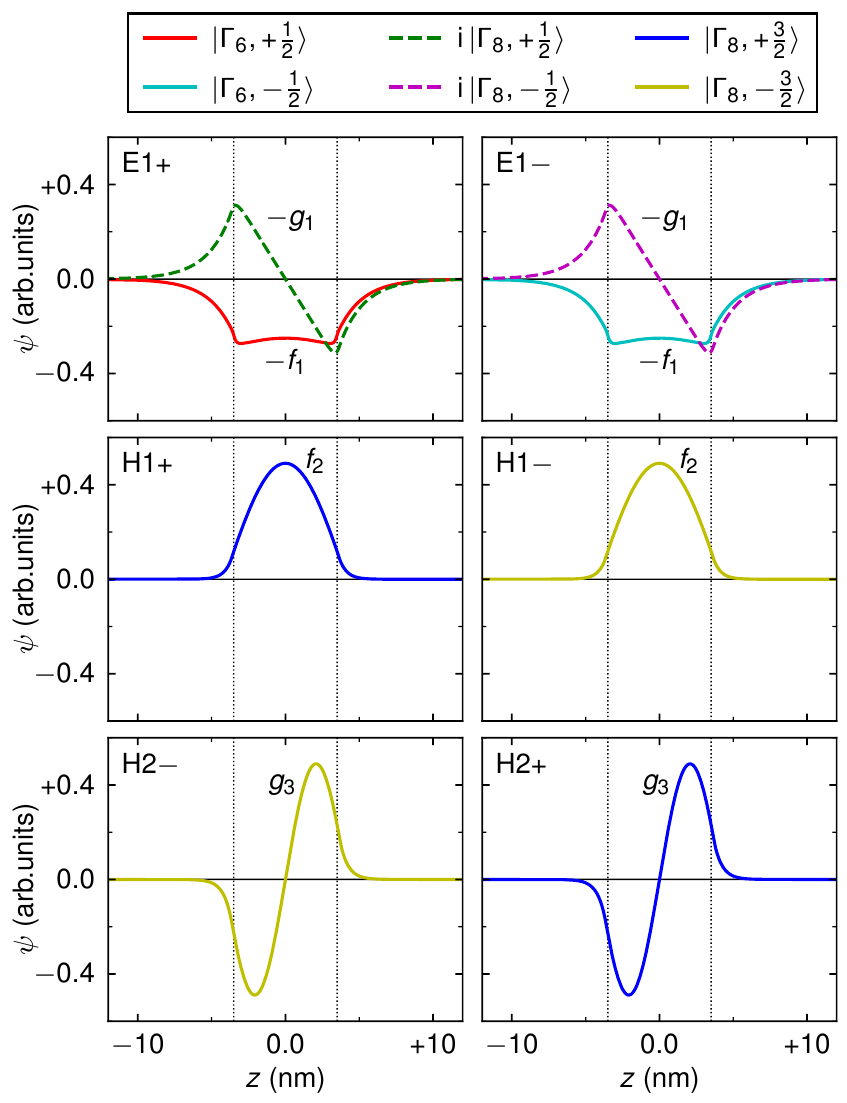}
  \caption{Wave functions for the subbands $\ket{\mathrm{E}1,\pm}$, $\ket{\mathrm{H}1,\pm}$, and $\ket{\mathrm{H}2,\pm}$ for a $7\nm$ HgTe quantum well at $(k_x,k_y)=(0,0)$. The colours distinguish the orbital contributions. Zero and negligible contributions (among which the $\ket{\Gamma_7,\pm\frac{1}{2}}$ orbitals) are not shown. Solid and dashed curves denote the real and imaginary parts, respectively. The vertical dotted lines indicate the interfaces between quantum well (HgTe) and barriers (Hg$_{1-x}$Cd$_{x}$Te). We label the envelope functions $f_i$ (even parity) and $g_i$ (odd parity).}
  \label{fig_wf}
\end{figure}


The particular grouping of the subband states given by \eqn\eqref{eqn_bhzblocks} can be readily understood from their expansion into orbitals and envelope functions \cite{RotheEA2010},
{\allowdisplaybreaks
\begin{align}
  \ket{\mathrm{E}1,\pm} &= - f_1(z)\ket{\Gamma_6,\pm\half} - \ii g_1(z)\ket{\Gamma_8,\pm\half},\nonumber\\
  \ket{\mathrm{H}1,\pm} &= f_2(z)\ket{\Gamma_8,\pm\threehalf},\label{eqn_envelopes}\\
  \ket{\mathrm{H}2,\pm} &= g_3(z)\ket{\Gamma_8,\pm\threehalf},\nonumber
\end{align}%
}%
where the small contributions from $\ket{\Gamma_7,\pm\half}$ have been neglected. The real functions $f_i$ are even, $g_i$ are odd under $z\to-z$, see Fig.~\ref{fig_wf}. The extended BHZ Hamiltonian thus groups states together which have either have identical spin and identical parity, or have opposite spin and opposite parity.

\subsection{Isoparity}

The fact that block-diagonal extensions of the BHZ model exist already hints at the possibility to find a similar block structure for the $\kdotp$ model, from which the former is derived. The blocks are distinguished by a conserved quantum number; the previous observations suggest that it must involves both angular momentum and parity under the transformation $z\to-z$. That is, the operator encoding this quantity should be the composition of the parity operation $P_z:z\mapsto -z,k_z\mapsto-k_z$ acting on spatial coordinates with a diagonal matrix $Q$ in orbital space,
\begin{equation}
 \isopz = P_z Q.
\end{equation}
We name this operator \emph{isoparity}, because of its nature being a combination of spatial parity with another degree of freedom (spin).


\begin{figure}
  \includegraphics{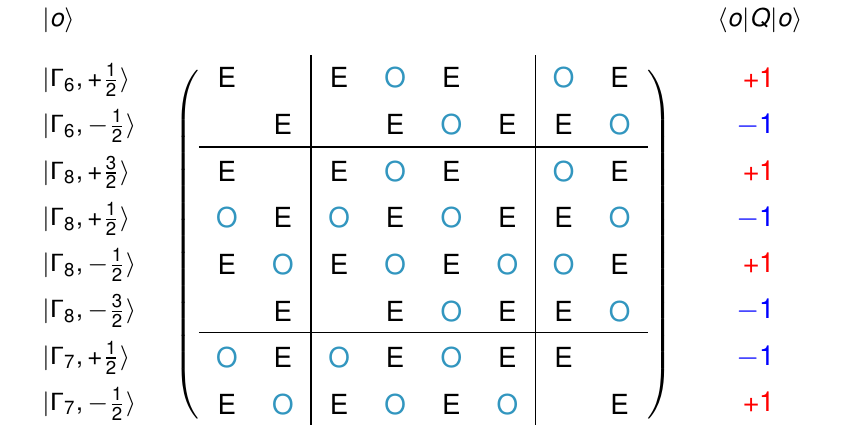}
  \caption{Parity structure of the Hamiltonian $H_k$ and construction of the isoparity matrix $Q$. The labels `E' and `O' denote even and odd parity, respectively, under $P_z:z\to-z,k_z\to-k_z$, of the particular entry of the Hamiltonian. Even-parity entries contain even powers of $k_z$ only (including constants), whereas odd-parity entries contain only odd powers. An empty position means that the matrix entry is identically zero. On the right hand side, we indicate the diagonal matrix values $\bramidket{o}{Q}{o}$ for each of the orbitals $\ket{o}$ in the basis.}
  \label{fig_ham_parity}
\end{figure}


The matrix $Q$ in orbital space contains eigenvalues $\pm 1$, such that opposite spins ($J_z=\pm m_J$) come with opposite signs. In order for the  operator $\isopz$ to be a conserved quantity, $[H,\isopz]=0$, it is required that $Q$ satisfies
\begin{equation}\label{eqn_evenodd}
  H^\mathrm{even} Q = Q H^\mathrm{even},
  \text{ and }
  H^\mathrm{odd} Q = -Q H^\mathrm{odd},
\end{equation}
where the Hamiltonian $H=H^\mathrm{even}+H^\mathrm{odd}$ is separated into parts that are even and odd under the mirror operation $P_z$, i.e., containing even and odd powers of $z$ or $k_z$, as indicated by the entries labelled `E' and `O' in Fig.~\ref{fig_ham_parity}. In other words, $Q$ commutes (anticommutes) with the even (odd) part of the Hamiltonian under $P_z$. Thus, we find
\begin{equation}\label{eqn_matq}
  Q = \diag(+1,-1,+1,-1,+1,-1,-1,+1).
\end{equation}
This matrix can simply be constructed by assigning the same diagonal value for all orbitals connected by even-parity matrix elements of the Hamiltonian, as shown in Fig.~\ref{fig_ham_parity}.
This top-down construction of isoparity provides a suitable practical definition that is a conserved quantity by construction.
In Sec.~\ref{sec_pointgroup}, we will provide a bottom-up derivation of this matrix from a fundamental perspective using representation theory of the crystallographic point group, that explains why it is conserved.

The conservation of isoparity in the $\kdotp$ model is illustrated in Fig.~\ref{fig_disp_colors}(c) with dispersion curves coloured according to the eigenstate expectation value $\avg{\isopz}$. Every eigenstate of the Hamiltonian is also an eigenstate of isoparity, with eigenvalue $\isopz=\pm 1$. The expectation values $\avg{\isopz}$ are consequently momentum independent: In the colour coding of Fig.~\ref{fig_disp_colors}(c), states are either uniformly red or uniformly blue, indicating that they have isoparity eigenvalue of $\isopz=+1$ and $-1$, respectively. In conclusion, the Hamiltonian admits a natural decomposition into two decoupled blocks defined by the isoparity eigenvalue.

Whereas we could also have chosen the opposite overall sign for $Q$, the definition of \eqn\eqref{eqn_matq} intuitively connects to the blocks of the $4$-band BHZ model. The basis states $\ket{\mathrm{E}1,\pm}$ and $\ket{\mathrm{H}1,\pm}$ are eigenstates of isoparity with eigenvalues $\pm1$,
\begin{equation}\label{eqn_isopz_e1h1}
  \isopz\ket{\mathrm{E}1,\pm}=\pm\ket{\mathrm{E}1,\pm},
  \text{ and }
  \isopz\ket{\mathrm{H}1,\pm}=\pm\ket{\mathrm{H}1,\pm}.
\end{equation}
The block quantum numbers defined by isoparity thus correspond to the $+$ and $-$ (or `up' and `down') labels for the blocks of the $4$-band BHZ model. For these four subband states, the isoparity eigenvalues happen to be equal to $\sgn(J_z)$, but that is not generically true: For other subbands, the eigenvalues may be opposite, e.g., $\isopz\ket{\mathrm{H}2,\pm}=\mp\ket{\mathrm{H}2,\pm}$.
Thus, the common interpretation of the two blocks as `spin blocks' is appropriate for the $4$-band BHZ model, but it breaks down for extended BHZ-type models.
This observation underlines the need for the labelling to be defined unambiguously. Here, we use the conventional notation $+,-$ in terms of $\sgn(J_z)=\pm1$. The other choice would be for $+,-$ to mean the isoparity eigenvalue. If both $\isopz=\pm1$ and $\sgn(J_z)=\pm1$ need to be considered, two distinct pairs of symbols need to be used, e.g., $\oplus,\ominus$ and $+,-$. The notation $\spinup,\spindn$ is discouraged.

\subsection{Conservation laws}

Let us verify invariance under isoparity $\isopz$, using the $\kdotp$ Hamiltonian split into orbital blocks $H_{pq}$, as in \eqn\eqref{eqn_ham_blocks}. These blocks need to satisfy $H^\mathrm{even}_{pq}Q_q = Q_pH^\mathrm{even}_{pq}$ and $H^\mathrm{odd}_{pq}Q_q = -Q_pH^\mathrm{odd}_{pq}$, where $Q_6=\diag(+1,-1)$, $Q_8=\diag(+1,-1,+1,-1)$, and $Q_7=\diag(-1,+1)$ are the parts of $Q$ [\eqn\eqref{eqn_matq}] belonging to each orbital sector. 
We test these conditions for a selection of blocks $H_{pq}$ explicitly, whereas the other conditions can be worked out along analogous lines. The blocks $H_{pq}$ are written in a formal manner in Appendix~\ref{sec_app_hamkdotp}, following the notation of Ref.~\cite{Winkler2003_book}.

The diagonal blocks $H_{66}$ and $H_{77}$ are proportional to the identity matrix in the orbital basis, and thus automatically satisfy $H^\mathrm{even}_{pp}Q_p = Q_pH^\mathrm{even}_{pp}$. The block $H_{88}$ [\eqn\eqref{eqn_h88}] contains a term proportional to the identity, as well as the two terms,
\begin{align}
  &\gamma_2 \left[(J_x^2-\tfrac{1}{3}J^2) k_x^2 + (J_y^2-\tfrac{1}{3}J^2) k_y^2 + (J_z^2-\tfrac{1}{3}J^2) k_z^2\right],\\
  &\gamma_3 \left[\{J_x,J_y\}\{k_x,k_y\}\!+\!\{J_x,J_z\}\{k_x,k_z\}\!+\!\{J_y,J_z\}\{k_y,k_z\}\right]\!,
\end{align}
where $J^2 = J_x^2+ J_y^2 +J_z^2$ and $\{A,B\}=AB+BA$ denotes the anticommutator.
The $\gamma_2$ term is purely even in $P_z$, and commutes with $Q_8$ by virtue of $[J_i^2,Q_8]=0$. For the $\gamma_3$ term, the even part is proportional to $\{k_x,k_y\}$, whereas $\{k_x,k_z\}$ and $\{k_y,k_z\}$ constitute the odd part. The identities $[\{J_x,J_y\},Q_8]=0$ and $\{\{J_i,J_z\},Q_8\}=0$ ($i=x,y$) then provide invariance of this term under isoparity. The invariance of the off-diagonal blocks [Eqs.~\eqref{eqn_h67}--\eqref{eqn_h78}] can be worked out in an analogous manner.
Combining these results, we thus establish that $H_k$ commutes with isoparity $\isopz$.

\subsection{External fields: magnetic, electric, and strain couplings}

In order to describe realistic experimental settings, the Hamiltonian describing the band structure acquires several terms related to external forces in addition to the purely momentum dependent $\kdotp$ Hamiltonian. Here, electric and magnetic fields are the perhaps the first ones coming into mind, but also strain (in response to external stress) is of a similar nature.

Let us illustrate the implications of the field being intrinsic or external by the example of the Zeeman coupling acting on the $\Gamma_6$ orbitals,
\begin{align}
  H_{\mathrm{Z},6} &=\vec{S}\cdot\vec{B}=S_xB_x+S_yB_y+S_zB_z\\
  &=\frac{1}{2}\begin{pmatrix}B_z&B_x-\ii B_y\\B_x+\ii B_y&-B_z\end{pmatrix},
\end{align}
where $\vec{S}=(S_x,S_y,S_z)$ is the spin operator and $\vec{B}$ the magnetic field.
If we assume the magnetic field to be \emph{intrinsic} (or if we consider the transformation to be a pure coordinate transformation), both $\vec{S}$ and $\vec{B}$ transform as axial vectors.
For example, under $P_z$,
\begin{align}
 (S_x,S_y,S_z)&\to(-S_x,-S_y,S_z),\nonumber\\
 (B_x,B_y,B_z)&\to(-B_x,-B_y,B_z),
\end{align}
so that $\vec{S}\cdot\vec{B}$ is invariant.
(The transformation of $\vec{S}$ follows from $S_i\to Q_6^\dagger S_i Q_6$.)

In contrast, if one considers the symmetry of the system subjected to an \emph{external} magnetic field, then the latter should be considered as an external parameter that is unaffected by the transformation (or, in other words, transforms trivially), while $\vec{S}$, being intrinsic to the system, still transforms as an axial vector. Then we have
\begin{align}
 (S_x,S_y,S_z)&\to(-S_x,-S_y,S_z),\nonumber\\
 (B_x,B_y,B_z)&\to(B_x,B_y,B_z),
\end{align}
such that
\begin{equation}\label{eqn_zeeman_transform}
  H_{\mathrm{Z},6}
  \to Q_6^\dagger H_{\mathrm{Z},6} Q_6
  = \frac{1}{2}\begin{pmatrix}B_z&-(B_x-\ii B_y)\\-(B_x+\ii B_y)&-B_z\end{pmatrix},
\end{equation}
i.e., $S_xB_x+S_yB_y+S_zB_z\to -S_xB_x-S_yB_y+S_zB_z$. The invariance of $\vec{S}\cdot\vec{B}$ is thus broken under this transformation if $B_x\ne 0$ or $B_y\ne 0$. Thus, the Zeeman-type term preserves isoparity if the field is perpendicular, $\vec{B}=(0,0,B_z)$.
For the Zeeman couplings in the other orbital sectors \cite{Winkler2003_book}, the same conclusion follows analogously.

These observations keep their validity if the coupling is of the form $\vec{S}\cdot\vec{m}$, where $\vec{m}$ is a magnetic moment with the same transformation properties as $\vec{B}$. In particular, in Hg$_{1-x}$Mn$_x$Te, the exchange coupling couples the carrier spins to the (average) magnetic moments of the Mn ions. The response $\vec{m}(\vec{B})$ is paramagnetic \cite{NovikEA2005}, so that $\vec{m}$ transforms like $\vec{B}$. Thus, the magnetic couplings of HgTe and Hg$_{1-x}$Mn$_x$Te have identical symmetry properties.

The magnetic field also modifies the  $\kdotp$ Hamiltonian following the Peierls substitution $\vec{k}\to\vec{k}+e\vec{A}/\hbar$, where $\vec{A}$ is the gauge potential, defined such that $\nabla\times\vec{A}=\vec{B}$. For isoparity to remain preserved, it is required that $\vec{A}$ transforms in a vector-like manner under $P_z:z\to-z$, i.e., $(A_x,A_y,A_z)\to(A_x,A_y,-A_z)$, similar to the transformation of momentum $\vec{k}$.
However, unlike momentum $\vec{k}$, the vector potential is defined with respect to the \emph{external} magnetic field $\vec{B}$: Fixing the gauge $\vec{A} = (-B_z y+B_y z, -B_x z, 0)$ and applying the transformation $P_z$, we find that $\vec{A}$ transforms to $\vec{A}'=(-B_z y-B_y z, B_x z, 0)$. The transformation $\vec{A}\to\vec{A}'$ is vector-like exactly if the in-plane components vanish, $B_x=B_y=0$. A nonzero out-of-plane component $B_z$ preserves isoparity. Importantly, these conditions are identical to that of the Zeeman-type couplings.
Consequently, isoparity remains a good quantum number in a perpendicular external magnetic field, but not if the in-plane components are nonzero.

An electric field (uniform in the in-plane coordinates $x,y$) can be modelled as an external potential $V(z)$ times the identity matrix $\idmat$ in the orbital basis. In view of condition \eqn\eqref{eqn_evenodd}, isoparity is preserved if
\begin{equation}
  [V^\mathrm{even}(z)\idmat,Q] = 0,
  \text{ and }
  \{V^\mathrm{odd}(z)\idmat,Q\} = 0,
\end{equation}
where the function $V(z)=V^\mathrm{even}(z)+V^\mathrm{odd}(z)$ is split into even and odd parts under $z\to-z$. Thus, isoparity remains a good quantum number provided the odd part $V^\mathrm{odd}(z)$ vanishes. It should be noted that the electric potential induced by a gate voltage has a nontrivial odd component, so that isoparity symmetry is broken. Consequently, the Rashba effect involves matrix elements off-diagonal in isoparity (block quantum number). This result is well known in the context of BHZ \cite{LiuEA2008PRL100,RotheEA2010}, but thus also applies to the more generic setting of $\kdotp$.

The deformation of the lattice as a result of matching the lattice constant to a substrate material is encoded as the rank-two symmetric strain tensor $\epsilon_{ij}$ \cite{PfeufferJeschke2000_thesis,NovikEA2005}. Strain can be considered as the response to stress, the pressure applied to bring the lattice out of its equilibrium geometry. These forces are applied externally, and hence stress and strain must be treated as external fields.
The strain Hamiltonian $\Hstr$ describes the couplings between the orbitals and the strain tensor $\epsilon_{ij}$. According to the formalism of Pikus and Bir, the possible couplings can be found from the $\kdotp$ Hamiltonian by the substitutions $\tfrac{1}{2}\{k_i,k_j\}\to\epsilon_{ij}$ \cite{BirPikus1974_book,PfeufferJeschke2000_thesis,NovikEA2005}. (See Appendix~\ref{sec_app_hamstrain} for details.) As an external field, $\epsilon_{ij}$ transforms trivially, in contrast to $\{k_i,k_j\}$: Whereas under parity in $z$, $\{k_x,k_z\}$ and $\{k_y,k_z\}$ are odd and $\{k_x,k_y\}$ and $k_i^2$ ($i=x,y,z$) are even, all components of $\epsilon_{ij}$ are even. From the invariance conditions of the $\kdotp$ Hamiltonian derived above, we thus find that strain preserves isoparity if $\epsilon_{xz} = \epsilon_{yz} = 0$. This is the case for a material lattice strained to a substrate, where these shear strain components are typically assumed to vanish.


\begin{figure}
  \includegraphics{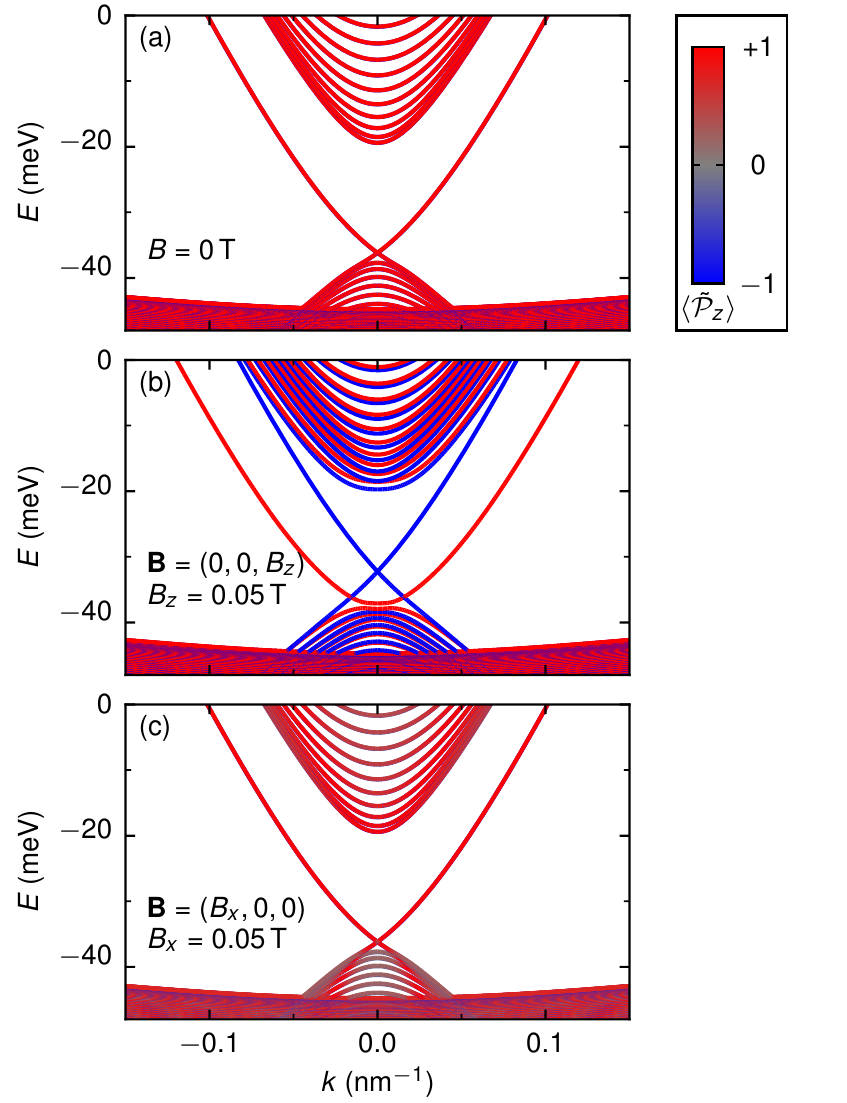}
  \caption{Dispersion for the $7\nm$ HgTe quantum well in a strip geometry, $500\nm$ wide. (a) At zero field, $B=0$, the eigenstates are degenerate, and can be separated into $\isopz=\pm1$ states, as indicated by the colour (see legend). (b) Dispersion for nonzero perpendicular field, $\vec{B}=B_z\hat{z}$ with $B_z = 0.05\,\mathrm{T}$. (c) Dispersion for nonzero in-plane field, $\vec{B}=B_x\hat{x}$ with $B_x = 0.05\,\mathrm{T}$.}
  \label{fig_magn}
\end{figure}


Figure~\ref{fig_magn} illustrates the effect of a magnetic field on isoparity on dispersions of the aforementioned quantum well in a strip geometry, $500\nm$ wide in the transversal direction $\hat{y}$  and translationally symmetric in the direction $\hat{x}$. This configuration models the devices used in experiments \cite{KonigEA2007,ShamimEA2020SciAdv,ShamimEA2021unpublMOTO}.
We have used the eight-orbital $\kdotp$ model from Ref.~\cite{NovikEA2005}. We include strain from lattice matching the HgTe crystal to the lattice constant of the Cd$_{0.96}$Zn$_{0.04}$Te substrate. The values of the coefficients are provided in Table~\ref{table_kdotpcoeff} in the Appendix. The magnetic field is modelled by the Peierls substitution $\vec{k}\to\vec{k}+e\vec{A}/\hbar$ with the vector potential $\vec{A} = (-B_z y, 0, 0)$ in Landau gauge.
For the confinement in $\hat{y}$ direction, we have used hard-wall boundary conditions. While a different choice of the boundary condition could affect the edge channel dispersion and the particular position of the Dirac point (the energy where the dispersions cross) \cite{Raichev2012,SkolasinskiEA2018PRB,Klipstein2020preprint}, it does not change the symmetry properties.

In absence of a magnetic field, the Hamiltonian eigenstates have block degeneracy; since $[H,\isopz]=0$, all eigenstates can be classified as $\isopz=\pm1$ [Fig.~\ref{fig_magn}(a)]. In a finite perpendicular magnetic field $\vec{B}=B_z\hat{z}$ the block degeneracy is lifted, but each Hamiltonian eigenstate is still an eigenstate of $\isopz$ [Fig.~\ref{fig_magn}(b)].  For an in-plane field, $\vec{B}=B_x\hat{x}$, $\isopz$ no longer commutes with the Hamiltonian: In Fig.~\ref{fig_magn}(c), we thus observe that the dispersion curves are coloured non-uniformly, especially at the top of the valence band.

With the `spin' of the edge channels being particularly important for spin transport, we can raise the question whether the existing interpretation of spin as $\sgn(J_z)=\pm1$ is adequate. In view of \eqn\eqref{eqn_isopz_e1h1}, the eigenvalues of $\sgn(J_z)$ and $\isopz$ coincide for the $\ket{\mathrm{E}1,\pm}$ and $\ket{\mathrm{H}1,\pm}$ subband states. 
For the configurations of Figs.~\ref{fig_magn}(a) and (b), the combined overlaps of each edge channel wave function with $\ket{\mathrm{E}1,\pm}$ and $\ket{\mathrm{H}1,\pm}$ is $>0.8$ for all momenta $\abs{k_x}\lesssim0.1\nm^{-1}$. Thus, in the regime where the edge states cross the bulk gap, $\sgn(J_z)$ may be used safely as a substitute for isoparity. Nonetheless, isoparity remains preferred in any configuration where it is an exactly conserved quantum number.

\subsection{Inversion symmetry breaking}
For the $\kdotp$ Hamiltonian, we have only included inversion symmetric terms so far. Since the zincblende lattice structure does not possess inversion symmetry, the Hamiltonian may contain inversion breaking terms. These are collectively known as bulk inversion asymmetry (BIA).

In order to demonstrate how BIA affects the block structure in the $\kdotp$ Hamiltonian defined by isoparity, let us study the linear BIA term of the $\Gamma_8$ block \cite{Winkler2003_book},
\begin{align}
  H_{\mathrm{BIA},C}&=C_k\left [\{J_x,J_y^2-J_z^2\}k_x + \{J_y,J_z^2-J_x^2\}k_y\right.\nonumber\\
  &\qquad\left.+ \{J_z,J_x^2-J_y^2\}k_z\right].
\end{align}
The $k_x$ and $k_y$ terms constitute the even part and the $k_z$ term is the odd part. Using the definitions of the matrices $J_i$, we calculate
\begin{align}
  \{\{J_x,J_y^2-J_z^2\},Q_8\}&=0\nonumber\\
  \{\{J_y,J_z^2-J_x^2\},Q_8\}&=0\\
  [\{J_z,J_x^2-J_y^2\},Q_8]&=0.\nonumber
\end{align}
For the even and odd parts of the Hamiltonian, this yields
\begin{equation}\label{eqn_evenodd_bia}
  H^\mathrm{even} Q = -Q H^\mathrm{even}
  \text{ and }
  H^\mathrm{odd} Q = Q H^\mathrm{odd},
\end{equation}
i.e., the even part anticommutes with $Q$ (here, $Q_8$) while the odd part commutes, the opposite from the behaviour of inversion symmetric terms, see \eqn\eqref{eqn_evenodd}. Thus, we find that $H_{\mathrm{BIA},C}$ anticommutes with isoparity,
\begin{equation}
  H_{\mathrm{BIA},C}\isopz = -\isopz H_{\mathrm{BIA},C}.
\end{equation}
The same property is valid for other bulk-inversion asymmetric terms; see, e.g., Ref.~\cite{Winkler2003_book}.

While the bulk-inversion antisymmetric terms are typically weaker than the symmetric terms \cite{OrlitaEA2011}, this is not generically valid for other inversion asymmetric  contributions. For example, structural inversion asymmetry (SIA) intrinsic to the geometry of the `type-II' heterostructure AlSb/InAs/GaSb/AlSb can be comparable to the topological gap \cite{LiuEA2008PRL100}. Electrostatic gating can also induce asymmetries well above $\sim10\meV$.

Even if isoparity is not a symmetry, it may be useful to write the Hamiltonian in a basis of isoparity eigenstates. It then takes the form
\begin{equation}\label{eqn_biaweak}
  H = \begin{pmatrix} H^+ & H_\mathrm{IA} \\ H_\mathrm{IA}^\dagger & H^- \end{pmatrix},
\end{equation}
where the blocks $H^\pm$ contain inversion symmetric matrix elements of $\isopz=\pm1$ states and $H_{IA}$ contains the inversion asymmetric terms. This structure is fully analogous to BHZ-type models in the basis $\ket{\mathrm{E}1,+}$, $\ket{\mathrm{H}1,+}$, $\ket{\mathrm{E}1,-}$, $\ket{\mathrm{H}1,-}$, where BIA and SIA terms appear in the off-diagonal block that couples $+$ and $-$ states \cite{KonigEA2008,LiuEA2008PRL100}. 
Only if the off-diagonal blocks are small, isoparity is \emph{weakly broken} and the inversion asymmetric terms can be considered perturbatively.

\section{Crystal symmetry: Point group analysis}
\label{sec_pointgroup}

In the previous section, we have derived the isoparity operator $\isopz$ in an \emph{a posteriori} manner, using invariance conditions with respect to a given Hamiltonian. In this section, we will derive a suitable notion of block quantum number from the fundamental principles of crystal symmetries. Since the Hamiltonian is necessarily compatible with crystal symmetry, statements about invariance follow automatically, by construction.

The symmetries of the $\kdotp$ model around $\vec{k}=\Gamma$ are described by one of the crystallographic point groups. The appropriate point group for unstrained bulk zincblende crystals is $T_d$ (tetrahedral symmetry). If the system is confined and/or strained in the $z$ direction, the point group is reduced to $D_{2d}$. Here, we consider the inversion symmetric approximation; adding inversion to these groups, we obtain $O_{h}$ and $D_{4h}$, respectively. In order to describe fermionic degrees of freedom, we need to extend these groups with the $2\pi$ rotation $\tilde{1}$. The resulting double covering groups are the \emph{double point groups} $\dpg{O}{h}$ and $\dpg{D}{4h}$ \cite{ElcoroEA2017}.

We consider the transformations of the orbitals in the basis in \eqn\eqref{eqn_kdotpbasis}. The four spin representations of the relevant double point group $\dpg{D}{4h}$ are $E_{1/2,g}$, $E_{3/2,g}$, $E_{1/2,u}$, and $E_{3/2,u}$, where the subscripts $1/2$ and $3/2$ denote the quantum number $m_J$ of the total angular momentum $J_z$ and the subscripts $g$ and $u$ denote the parity under spatial inversion $I:(x,y,z)\to(-x,-y,-z)$, being \emph{gerade} (even) and \emph{ungerade} (odd), respectively. This parity corresponds to the orbital angular momentum $L$ being an even or odd integer. We follow the notation used in the tables of Ref.~\cite{AltmannHerzig1994}, that is a slight modification of the common `Mulliken notation', used in Refs.~\cite{Mulliken1933,BradleyCracknell1972,ElcoroEA2017} for example. The aforementioned four spin representations are written as $\bar{E}_{1g}$, $\bar{E}_{2g}$, $\bar{E}_{1u}$, and $\bar{E}_{2u}$, respectively, in the Mulliken notation.

In Table~\ref{table_d4hrep}, we list the orbitals of the $\kdotp$ together with the angular momentum values $L$, $S$, and $J$ (orbital, spin, and total, respectively) and $m_J$. From these values, we identify the appropriate spin representation. We subsequently provide the representation matrices $r(g)$ for several relevant group elements $g$: the mirrors $m_i$ along the axes $i=x,y,z$.
These representation matrices encode linear transformation on the two-dimensional vector spaces with basis $\ket{\Gamma_p,+m_J}, \ket{\Gamma_p,-m_J}$ ($m_J>0$), where the positive $m_J$ is the first component. We have followed the conventions of the point group tables of Ref.~\cite{AltmannHerzig1994}, which we also recommend as a reference for additional information about this and other point groups. We remark that $\ket{\Gamma_7,\pm\half}$ and $\ket{\Gamma_8,\pm\half}$ transform both according to representation $E_{1/2,u}$, but the representation matrices are different, related by a unitary transformation.

\begin{table}
 \begin{tabular}{c|cccc}
  subspace & $\ket{\Gamma_6,\pm\half}$ & $\ket{\Gamma_7,\pm\half}$ & $\ket{\Gamma_8,\pm\half}$ & $\ket{\Gamma_8,\pm\threehalf}$ \\
  \hline
  $L$     & $0$     & $1$     & $1$     & $1$ \\
  $S$     & $\half$ & $\half$ & $\half$ & $\half$ \\
  $J$     & $\half$ & $\half$ & $\threehalf$ & $\threehalf$ \\
  $m_J$   & $\pm\half$ & $\pm\half$ & $\pm\half$ & $\pm\threehalf$\\
  \hline
  $\dpg{D}{4h}$ rep. & $E_{1/2,g}$ & $E_{1/2,u}$ & $E_{1/2,u}$ & $E_{3/2,u}$\\
                     & $\bar{E}_{1g}$ & $\bar{E}_{1u}$  & $\bar{E}_{1u}$  & $\bar{E}_{2u}$\\
  \hline
  $m_z$
  & $\begin{pmatrix}-\ii&0\\0& \ii\end{pmatrix}$
  & $\begin{pmatrix} \ii&0\\0&-\ii\end{pmatrix}$
  & $\begin{pmatrix} \ii&0\\0&-\ii\end{pmatrix}$
  & $\begin{pmatrix}-\ii&0\\0& \ii\end{pmatrix}$\\
  $m_x$
  & $\begin{pmatrix}0&-\ii\\-\ii&0\end{pmatrix}$
  & $\begin{pmatrix}0& \ii\\ \ii&0\end{pmatrix}$
  & $\begin{pmatrix}0&-\ii\\-\ii&0\end{pmatrix}$
  & $\begin{pmatrix}0&-\ii\\-\ii&0\end{pmatrix}$\\
  $m_y$
  & $\begin{pmatrix}0&-1\\ 1&0\end{pmatrix}$
  & $\begin{pmatrix}0& 1\\-1&0\end{pmatrix}$
  & $\begin{pmatrix}0&-1\\ 1&0\end{pmatrix}$
  & $\begin{pmatrix}0& 1\\-1&0\end{pmatrix}$\\
  \hline
  TR $\theta$
  & $\begin{pmatrix}0&-1\\ 1&0\end{pmatrix}$
  & $\begin{pmatrix}0&-1\\ 1&0\end{pmatrix}$
  & $\begin{pmatrix}0& 1\\-1&0\end{pmatrix}$
  & $\begin{pmatrix}0&-1\\ 1&0\end{pmatrix}$
 \end{tabular}
 \caption{For each of the two-dimensional spinor subspaces in the top row, which together span the $8$-orbital basis of the $\kdotp$ model, the table provides: the angular momentum quantum numbers $L$, $S$, $J$, and $m_J$; the representation of the double point group $\dpg{D}{4h}$ (two common notations); representation matrices $r(g)$ in for a selected subset of group elements $g$; the matrix $\theta$ of time reversal $T=\theta K$. We follow the conventions of the point group tables of Ref.~\cite{AltmannHerzig1994}. }
 \label{table_d4hrep}
\end{table}

The representation $R$ for the $8$-orbital $\kdotp$ basis defined by \eqn\eqref{eqn_kdotpbasis}, can simply be composed from those of the two-dimensional subspaces listed in Table~\ref{table_d4hrep}. Thus, the matrix $R(m_z)$ for $m_z$ reads
\begin{equation}\label{eqn_mz_8o}
  R(m_z) = \diag(-\ii,+\ii,-\ii,+\ii,-\ii,+\ii,+\ii,-\ii).
\end{equation}
This matrix is identical to the matrix $Q$ [\eqn\eqref{eqn_matq}] for isoparity up to a factor, $R(m_z)=-\ii Q$. In view of the action of $m_z$ on spatial quantities being given by $P_z$, we conclude that isoparity $\isopz$ is nothing else than the mirror operation in $z$ up to the factor $\ii$. The two factors $P_z$ and $Q$ have a common origin, as the actions of $m_z$ on the spatial and orbital degrees of freedom, respectively.
This observation establishes why isoparity is conserved: The symmetry is of crystallographic origin. The transformations of the spinors follow naturally from the spin representations of the point group.
By analogy, the matrices for $R(m_x)$ and $R(m_y)$ may be used to construct `in-plane isoparity'.

\section{Time reversal}
\label{sec_timereversal}

Time reversal $T$ acts differently on position and momentum, $\vec{x}\to\vec{x}$ and $\vec{k}\to-\vec{k}$ respectively, as opposed to crystal symmetries where position and momentum transform in the same way. Consequently, time reversal also reverses angular momentum, including spin. 
Time reversal is required to be antiunitary in order the preserve the canonical quantization rules.
In order to achieve these properties, time reversal is typically realized as the combination $T=\theta K$ of a matrix $\theta$ in orbital space and complex conjugation, denoted $K$. For states with angular momentum $J$, this operator is most commonly represented by $\theta = \exp(-\ii\pi \hat{J}_y)$, itself a real matrix. The formal definition of time reversal requires the point group to be extended with time-reversed symmetry elements \cite{LandauLifshitz1985book_StatPhys,TavgerZaitsev1956JETP}, but for the present case of crystals without magnetic ordering, it suffices to do the algebra by inserting complex conjugation $K$ at the appropriate positions.

We specify the form of $\theta$ in all orbital subspaces of the $\kdotp$ basis in Table~\ref{table_d4hrep}. In all four two-dimensional subspaces, $\theta = r(2_y) = \pm r(m_y)$, with $+$ and $-$ for \emph{gerade} ($L=0$) and \emph{ungerade} ($L=1$) representations, where $2_y$ denotes the $\pi$ rotation around the $y$ axis.
From $r(2_y)r(m_z) = -r(m_z)r(2_y)$
and the fact that $r(m_z)$ is purely imaginary, it follows that time reversal commutes with $m_z$,
\begin{align}
  T r(m_z)
  &= \theta K r(m_z) = -\theta r(m_z) K = r(m_z) \theta K\nonumber\\
  &= r(m_z) T. \label{eqn_commutator_tr_mz}
\end{align}
Consequently, for a state belonging to one block, i.e., an eigenstate $\ket{\Psi}$ of $r(m_z)$ with $r(m_z)\ket{\Psi} = \pm\ii\ket{\Psi}$, we find that $r(m_z) T\ket{\Psi}=\mp\ii T\ket{\Psi}$, due to the complex conjugation in $T$.

Analogously, isoparity $\isopz$ \emph{anticommutes} with time reversal $T$, so that for $\ket{\Phi}$ with $\isopz\ket{\Phi}=\pm\ket{\Phi}$, we have $\isopz T\ket{\Phi} = \mp T\ket{\Phi}$.
Thus, the two members $\ket{\Psi}$ and $T\ket{\Psi}$ of a Kramers pair have opposite block quantum numbers.
This property allows the interpretation of the two blocks as being Kramers partners. This fundamental idea, well known in the context the BHZ model, is also valid in the more generic $\kdotp$ framework.
As such, the division of the $\kdotp$ Hamiltonian into isoparity subspaces $\isopz=\pm1$ can be considered as a full generalization of the blocks in the BHZ model.

\section{Magnetotransport observables}
\label{sec_cond_isocond}

While symmetry forces the transversal conductivity $\sigma_{xy}$ to vanish at zero magnetic field, the salient feature of the quantum spin Hall effect is the presence of helical edge channels, that carry a spin Hall current. These edge channels come as a counterpropagating pair of Kramers partners. Their contributions $\sigma_{xy}^{\pm}$ to the Hall conductivity cancel out, $\sigma_{xy}=\sigma_{xy}^{+}+\sigma_{xy}^{-}=0$, as to preserve the symmetry relation $\sigma_{xy}=0$. On the other hand, the difference $\tilde{\sigma}_{xy}=\sigma_{xy}^{+}-\sigma_{xy}^{-}$ is not required to vanish. The two partners being related by time reversal, they can be interpreted as two states with opposite spin, and one can thus call $\tilde{\sigma}_{xy}$ the spin Hall conductivity \cite{BernevigZhang2006}.

With the interpretation of the blocks $\isopz=\pm1$ as Kramers partners, we can utilize isoparity to generalize the concept of spin conductance to the $\kdotp$ framework: The difference between the conductivity tensor associated to either block, is simply a composition of the conductivity tensor $\sigma$ with the isoparity operator $\isopz$,
\begin{equation}\label{eqn_spinconductivitydef}
 \tilde\sigma_{ij} = \sigma_{ij} \isopz.
\end{equation}
While ordinary conductivity $\sigma$ satisfies Ohm's law $\vec{j} = \sigma \vec{E}$, spin conductivity satisfies the analogous equation $\tildevecj=\tilde{\sigma} E$, involving the \emph{spin current} $\tildevecj=\vec{j}\isopz$, that can be viewed in terms of two types of carriers with opposite charge depending on the isoparity eigenvalue.

The transformation rules of $\sigma$ and $\tilde{\sigma}$ under mirror symmetries readily follow from the group theoretical analysis of Sec.~\ref{sec_pointgroup}: Isoparity $\isopz = \ii m_z$ commutes with $m_z$ and anticommutes with $m_x$ and $m_y$. Under the latter two transformations, spin conductivity $\tilde\sigma$ obtains an opposite sign compared to charge conductivity $\sigma$, so that
\begin{equation}\label{eqn_tildesigmaxy_magnz}
 \sigma_{xy}(-B_z) = -\sigma_{xy}(B_z),\qquad
 \tilde\sigma_{xy}(-B_z) = \tilde\sigma_{xy}(B_z).
\end{equation}
Unlike the ordinary Hall conductivity $\sigma_{xy}$, the spin Hall conductivity $\tilde\sigma_{xy}$ is not required to vanish at zero magnetic field. In other words, spin Hall conductivity may be nonzero at zero magnetic field, which makes the existence of helical counterpropagating edge channels possible.

The transformation of transport tensors $L_{ij}$ under time reversal $T$ is subject to the principle of Onsager's relation \cite{Grimmer1993,Grimmer2017},
\begin{equation}\label{eqn_onsager}
  L_{ij}(\vec{S},\vec{B}) = L_{ji}(-\vec{S},-\vec{B}),
\end{equation}
where $\vec{S}$ encodes the spin configuration of the system. The key property of $\vec{S}$ is that $\vec{S}\to-\vec{S}$ under time reversal. Here, isoparity plays this role. The conductivity contributions $\sigma_{ij}^{\pm}$ from the isoparity blocks $\isopz=\pm1$ thus satisfy
\begin{equation}
  \sigma_{ij}^{\pm}(\vec{B}) = \sigma_{ji}^{\mp}(-\vec{B}).
\end{equation}
(We note that in Chern insulator models with a single spin or isoparity block, see Refs.~\cite{BottcherEA2019,BottcherEA2020PRB} for example, there is no such relation.)
For the ordinary conductivity $\sigma_{ij}=\sigma_{ij}^+ + \sigma_{ij}^-$, the symmetric $\sigma^{s}_{ij}=\frac{1}{2}(\sigma_{ij}+\sigma_{ji})$ and antisymmetric $\sigma^{a}_{ij}=\frac{1}{2}(\sigma_{ij}-\sigma_{ji})$ parts are even and odd under time reversal, respectively,
\begin{equation}\label{eqn_sigmaij_magn}
 \sigma_{ij}^s(-\vec{B})=\sigma_{ij}^s(\vec{B}),\quad
 \sigma_{ij}^a(-\vec{B})=-\sigma_{ij}^a(\vec{B}),
\end{equation}
In contrast, spin conductivity $\tilde\sigma_{ij}=\sigma_{ij}^+ - \sigma_{ij}^-$  obtains opposite signs compared to $\sigma_{ij}$, 
\begin{equation}\label{eqn_tildesigmaij_magn}
 \tilde\sigma_{ij}^s(-\vec{B})=-\tilde\sigma_{ij}^s(\vec{B}),\quad
 \tilde\sigma_{ij}^a(-\vec{B})=\tilde\sigma_{ij}^a(\vec{B}).
\end{equation}
The opposite signs originate in the isoparity component $\isopz$ [see \eqn\eqref{eqn_spinconductivitydef}], which anticommutes with time reversal.

For $\vec{B}=0$, Eqs.~\eqref{eqn_sigmaij_magn} and \eqref{eqn_tildesigmaij_magn} thus express the idea that time reversal forces Hall conductivity $\sigma_{xy}$ to vanish while spin Hall conductivity $\tilde\sigma_{xy}$ may be finite.
For finite magnetic fields $\vec{B}\ne0$, Onsager's relation merely connects the two opposite isoparity blocks at $\vec{B}$ and $-\vec{B}$.
The values $\sigma^+_{xy}(\vec{B})$ and $\sigma^-_{xy}(\vec{B})$ are independent and may be finite, which permits the quantum Hall effect as well as the quantum anomalous Hall effect \cite{LiuEA2008PRL101}.
We conclude from this discussion that the concepts and properties of these transport tensors known from the BHZ formalism extend to $\kdotp$.
The key property of isoparity that accommodates this generalization is the idea that the isoparity sectors are Kramers partners.

\section{Conclusion and discussion}
\label{sec_conclusion}

The term \emph{quantum spin Hall effect} is more subtle than it appears at first sight: The effect pairs a zero total Hall conductivity with a nonzero spin Hall conductivity, being the sum and difference, respectively, of contributions from two `spin' states. Here, we have demonstrated that these two `spin' states are neither proper spin $S_z$ nor total angular momentum $J_z$, but the eigenvalues of the mirror transformation $m_z$ or the equivalent observable that we dubbed \emph{isoparity} $\isopz$.
The block structure of the Hamiltonian is defined by the quantum numbers defined by $m_z$ or $\isopz$ (values are $\pm\ii$ or $\pm1$, respectively).
The conservation of these quantities naturally follows from crystal symmetries.
The key property that makes isoparity `behave like spin' is the fact that it anticommutes with time reversal. As time reversal maps the $\isopz=+1$ subspace to the $\isopz=-1$ subspace, they constitute Kramers partners, analogous to the same idea in the BHZ model framework.
Thus, by analogy, spin conductivity is defined as the transport tensor $\tilde\sigma_{ij}=\sigma_{ij}\isopz$, defined by combining ordinary conductivity tensor $\sigma_{ij}$ with isoparity.
These analogies establish the isoparity subspaces as a complete generalization of the notion of the block structure of the BHZ model.
In conclusion, while `spin' in `quantum spin Hall' should be more accurately understood as isoparity, the analogy justifies the colloquial usage of the word `spin'.

The conservation of isoparity requires $m_z$ to be a crystal symmetry, which is however not contained in the point group $\dpg{T}{d}$ of (unstrained bulk) zincblende materials. The block structure can thus be defined by isoparity only in the inversion symmetric approximation: By neglecting inversion asymmetric terms, the point group is enlarged to $\dpg{O}{h}$, or one of its subgroups that contain inversion.
While actual topological insulator devices are often not symmetric under $z\to-z$, modelling of the band structures in the inversion symmetric approximation (including BHZ-like models) is ubiquitous in literature.
Even in absence of this symmetry, a basis of isoparity eigenstates may be chosen. This will lead to a block structure of the Hamiltonian with isoparity-preserving terms in the diagonal blocks, and the isoparity breaking terms in the off-diagonal blocks, again in full analogy to the BHZ model. A perturbative treatment of the symmetry breaking terms is possible if they are sufficiently small.

In practice, the isoparity operator can be applied in $\kdotp$ numerics in order to lift degeneracy between the blocks, by addition of a small term $\epsilon \isopz$ to the Hamiltonian. Whereas one could successfully lift degeneracy  with diagonal operators like $\epsilon J_z$ or $\epsilon\,\sgn(J_z)$, these have the disadvantage that they do not constitute conserved quantum numbers. In fact, we have demonstrated that the expectation values $\avg{J_z}$ can change signs at larger momentum values, which leads to accidental crossings, that complicates the study of the properties of single bands. These issues are avoided if the lifting term is given by the conserved quantity $\isopz$. Despite the operator $\isopz$ being non-diagonal, it can be implemented conveniently and efficiently in sparse-matrix form.



\acknowledgments
We thank J.~B\"ottcher, G.~Sangiovanni, and D.~Di~Sante for useful discussions and feedback on the draft version of this article.
We acknowledge financial support from
the Deutsche Forschungsgemeinschaft (DFG, German Research Foundation) in the project SFB 1170 (Project ID 258499086),
from the W\"urzburg--Dresden Cluster of Excellence on Complexity and Topology in Quantum Matter (EXC 2147, project ID 39085490),
and the `Institute for Topological Insulators', from the Free State of Bavaria.
The large-scale calculations of Fig.~\ref{fig_magn} were performed at the Julia-HPC high-performance-computing cluster at the University of W\"urzburg.


\appendix
\section{Inversion symmetric \kdotp\ Hamiltonian}
\label{sec_app_hamkdotp}

The inversion symmetric $\kdotp$ Hamiltonian $H_k$ is split into orbital blocks $H_{pq}$ ($p,q=6,7,8$) according to \eqn\eqref{eqn_ham_blocks}. (We drop the subscript $k$ in the orbital block notation.) We provide an abstract notation in terms of angular momentum matrices \cite{Winkler2003_book}. The latter notation is especially convenient for testing invariance. The explicit matrix forms can be found in Refs.~\cite{Winkler2003_book,PfeufferJeschke2000_thesis,NovikEA2005} for example, or from substitution of the angular momentum matrices.

For the diagonal blocks, the relevant angular momentum matrices are the $2\times 2$ Pauli matrices
\begin{equation}
 \sigma_x = \begin{pmatrix}0&1\\1&0\end{pmatrix},\qquad
 \sigma_y = \begin{pmatrix}0&-\ii \\ \ii&0\end{pmatrix},\qquad
 \sigma_z = \begin{pmatrix}1&0\\0&-1\end{pmatrix},
\end{equation}
and the $4\times4$ $J=\threehalf$ angular momentum matrices
\begin{align}
 J_x &= \begin{pmatrix}0&\theta&0&0\\\theta&0&1&0\\0&1&0&\theta\\0&0&\theta&0\end{pmatrix},&
 J_y &= \ii \begin{pmatrix}0&-\theta&0&0\\\theta&0&-1&0\\0&1&0&-\theta\\0&0&\theta&0\end{pmatrix},\nonumber\\
 J_z &= \begin{pmatrix}\threehalf&0&0&0\\0&\half&0&0\\0&0&-\half&0\\0&0&0&-\threehalf\end{pmatrix},
\end{align}
with $\theta=\half\sqrt{3}$, as well as the respective identity matrices. The diagonal blocks of $H_k$ are
\begin{align}
  H_{66} &= \left(E_6 +\frac{\hbar^2}{2m'}k^2\right)\idmat_{2\times2},\label{eqn_h66}\\
  H_{77} &= \left(E_7 -\frac{\hbar^2}{2m_0}\gamma_1 k^2\right)\idmat_{2\times2},\label{eqn_h77}
\end{align}
and
\begin{align}
 H_{88} &= \left(E_8 -\frac{\hbar^2}{2m_0}\gamma_1 k^2\right)\idmat_{4\times4}
         + \frac{\hbar^2}{2m_0}2\gamma_2 \left[(J_x^2-\tfrac{1}{3}J^2)k_x^2\right.\nonumber\\
          &\qquad\left. {}+ (J_y^2-\tfrac{1}{3}J^2) k_y^2 + (J_z^2-\tfrac{1}{3}J^2) k_z^2\right]\nonumber\\
          &\qquad {}+\frac{\hbar^2}{2m_0}\gamma_3 \left[\{J_x,J_y\}\{k_x,k_y\} + \{J_x,J_z\}\{k_x,k_z\}\right.\nonumber\\
          &\qquad\left. {}+ \{J_y,J_z\}\{k_y,k_z\}\right],\label{eqn_h88}
\end{align}
where $J^2 = J_x^2+ J_y^2 +J_z^2$ and $k^2=k_x^2+k_y^2+k_z^2$, $E_p$ ($p=6,7,8$) are the energies of the band edges at $k=0$, $m_0$ is the electron mass, $m'$ is the $\Gamma_6$ effective mass, and $\gamma_{1,2,3}$ are the Luttinger parameters \cite{PfeufferJeschke2000_thesis,Winkler2003_book}. The curly brackets denote the anticommutator, $\{A,B\}=AB+BA$. The off-diagonal blocks are
{\allowdisplaybreaks
\begin{align}
  H_{67} &= -\tfrac{1}{3}\sqrt{3}P(\sigma_x k_x +\sigma_y k_y + \sigma_z k_z),\label{eqn_h67}\\
  H_{68} &= -\sqrt{3}P(T_xk_x+T_yk_y+T_zk_z),\label{eqn_h68}\\
  H_{78} &= \frac{\hbar^2}{2m_0}6\gamma_2 \left(T_{xx}k_x^2 + T_{yy}k_y^2 + T_{zz}k_z^2\right)\label{eqn_h78}
    \\
    &{}+\frac{\hbar^2}{2m_0}6\gamma_3 \left(T_{xy}\{k_x,k_y\}\!+\!T_{yz}\{k_y,k_z\}\!+\!T_{zx}\{k_z,k_x\}\right).\nonumber
\end{align}%
}%
in terms of the $2\times 4$ angular momentum matrices $T_{i}$ and $T_{ij}$ ($i,j=1,2,3$; see Ref.~\cite{Winkler2003_book}), the Kane matrix element $P=-(\hbar/m_0)\bramidket{S}{p_x}{X}$ \cite{Kane1957} and the Luttinger parameters $\gamma_{2,3}$.


\begin{table*}
 \begin{tabular}{l|rrrrrrrr|rrrrrr}
  & \multicolumn{1}{c}{$E_6$} & \multicolumn{1}{c}{$E_7$} &  \multicolumn{1}{c}{$E_8$} & \multicolumn{1}{c}{$m'/m_0$} & \multicolumn{1}{c}{$P$}
  & \multicolumn{1}{c}{$\gamma_1$} & \multicolumn{1}{c}{$\gamma_2$} & \multicolumn{1}{c|}{$\gamma_3$}
  &  \multicolumn{1}{c}{$C$} & \multicolumn{1}{c}{$a$} &  \multicolumn{1}{c}{$b$} &  \multicolumn{1}{c}{$d$} & \multicolumn{1}{c}{$\epsilon_{\parallel}$} & \multicolumn{1}{c}{$\epsilon_{\perp}$}\\
  & \multicolumn{1}{c}{$\mathrm{meV}$} & \multicolumn{1}{c}{$\mathrm{meV}$} & \multicolumn{1}{c}{$\mathrm{meV}$} & & \multicolumn{1}{c}{$\mathrm{meV}\,\mathrm{nm}$}
  & & &
  & \multicolumn{1}{c}{$\mathrm{eV}$} & \multicolumn{1}{c}{$\mathrm{eV}$} & \multicolumn{1}{c}{$\mathrm{eV}$} & \multicolumn{1}{c}{$\mathrm{eV}$} & \multicolumn{1}{c}{$10^{-3}$} & \multicolumn{1}{c}{$10^{-3}$}\\
  \hline
  HgTe    & $-303$ & $-1080$ &    $0$ &      $1$ & $846$
  & $4.1$ & $0.50$ &  $1.30$
  & $-3.83$ & $0.00$ & $-1.50$ & $-2.08$ & $0.76$ & $-1.04$\\
  Hg$_{0.32}$Cd$_{0.68}$Te  &  $587$ & $-1343$ & $-379$ &   $1.14$ & $846$
  & $2.3$ & $-0.02$ & $0.43$
  & $-3.99$ & $-0.48$ & $-1.26$ & $-2.44$ & $-1.11$ & $1.54$\\
 \end{tabular}
 \caption{Coefficients of the inversion-symmetric $\kdotp$ Hamiltonian $H_k$ and the strain Hamiltonian $\Hstr$. The values are adapted from Refs.~\cite{PfeufferJeschke2000_thesis,NovikEA2005}, with values for HgTe provided as is, and those for Hg$_{0.32}$Cd$_{0.68}$Te obtained from interpolation between HgTe and CdTe.}
 \label{table_kdotpcoeff}
\end{table*}


The dispersions and wave functions of Figs.~\ref{fig_disp_colors} and \ref{fig_wf} have been calculated in a layer stack geometry with a $7\nm$ HgTe quantum well between barriers of Hg$_{1-x}$Cd$_{x}$Te with $x=0.68$. The thickness of the barriers in the simulation is $10\nm$, sufficient for the wave functions to decay.
In this geometry, there is no translational symmetry in the $z$ direction. The appropriate Hamiltonian in this geometry has the form $H(k_x,k_y,z)$ and is obtained from the bulk Hamiltonian $H(k_x,k_y,k_z)$ by the substitution $k_z\to -\ii \partial_z$. The coefficients of the $\kdotp$ Hamiltonian being material dependent, they are realized as piecewise constant functions of $z$ with smooth interpolation near the interface. The values for HgTe \cite{NovikEA2005} and Hg$_{1-x}$Cd$_{x}$Te at $x=0.68$ are summarized in Table~\ref{table_kdotpcoeff}.

The $z$ coordinate is discretized in steps of $0.25\nm$. The resulting Hamiltonian $H(k_x,k_y)$ has dimensions $N\times N$ with $N=8n_z$, where $n_z$ is the number of discrete coordinates in $z$ direction. The dispersions $E_n(k_x,k_y)$ are obtained by numerical diagonalization of this matrix. The wave functions and expectation values ($\avg{S_z}$, $\avg{J_z}$, etc.) are extracted from the eigenstates.
In the strip calculation of Fig.~\ref{fig_magn}, the Hamiltonian $H(k_x,y,z)$ is discretized additionally in the $y$ direction.  This leads to a Hamiltonian matrix $H(k_x)$ of dimension $N\times N$ with $N=8n_zn_y$, where $n_y$ is the number of discrete coordinates in $y$ direction. In view of the large matrix sizes, diagonalization has been performed with sparse matrix methods at a high-performance-computing infrastructure with a high degree of parallelization.

\section{Strain Hamiltonian}
\label{sec_app_hamstrain}
The uniaxial strain induced by lattice matching of the materials to the substrate is modelled by the Pikus-Bir strain Hamiltonian $\Hstr$ in terms of the strain tensor $\epsilon_{ij}$ ($i,j=x,y,z$) \cite{BirPikus1974_book}. In the Pikus-Bir formalism, the strain Hamiltonian is obtained by substitution of quadratic momenta $\frac{1}{2}\{k_i,k_j\}\to\epsilon_{ij}$ from the $\kdotp$ Hamiltonian $H_k$. 
The coefficients of the terms are the deformation potentials $C$, $a$, $b$, and $d$ \cite{PfeufferJeschke2000_thesis,NovikEA2005}. (Other notations are common, e.g., $C_1=C$, $D_d = a$, $D_u=-\frac{3}{2}b$, and $D'_u =-\frac{1}{2}\sqrt{3}d$ in Ref.~\cite{Winkler2003_book}.) The values provided in Table~\ref{table_kdotpcoeff} are extracted from Ref.~\cite{PfeufferJeschke2000_thesis}.

The substitution from $H_k$ yields the diagonal blocks
\begin{align}
 \Hstr_{66} &= C \tr \epsilon\, \idmat_{2\times2},\label{eqn_hstr66}\\
 \Hstr_{77} &= a \tr \epsilon\, \idmat_{2\times2},\label{eqn_hstr77}\\
 \Hstr_{88} &= a \tr \epsilon\, \idmat_{4\times4}
          -b \left[(J_x^2-\tfrac{1}{3}J^2)\epsilon_{xx}\right.\label{eqn_hstr88}\\
          &\qquad\left. {}+ (J_y^2-\tfrac{1}{3}J^2) \epsilon_{yy} + (J_z^2-\tfrac{1}{3}J^2) \epsilon_{zz}\right]\nonumber\\
          &\quad{}-\frac{d}{\sqrt{3}} \left[\{J_x,J_y\}\epsilon_{xy} + \{J_x,J_z\}\epsilon_{xz}+ \{J_y,J_z\}\epsilon_{yz}\right],\nonumber
\end{align}
with $\tr \epsilon = \epsilon_{xx} + \epsilon_{yy} + \epsilon_{zz}$.
Furthermore, we have the off-diagonal block
\begin{align}
  \Hstr_{78} &= -3b \left(T_{xx}\epsilon_{xx} + T_{yy}\epsilon_{yy} + T_{zz}\epsilon_{zz}\right)\label{eqn_hstr78}
    \\
    &{}\quad{}-\sqrt{3}d \left(T_{xy}\epsilon_{xy}+T_{yz}\epsilon_{yz}+T_{zx}\epsilon_{zx}\right).\nonumber
\end{align}%
The off-diagonal blocks $\Hstr_{68}$ and $\Hstr_{67}$ vanish as the corresponding blocks of $H_k$ do not contain terms of quadratic or higher order in $k$ (under the assumption of inversion symmetry).

The strain tensor $\epsilon_{ij}$ is determined by matching the in-plane lattice constants of the quantum-well and barrier layers are matched to the lattice constant $0.6467\nm$ of the Cd$_{0.96}$Zn$_{0.04}$Te substrate.
The equilibrium lattice constants for HgTe (quantum well) and Hg$_{0.32}$Cd$_{0.68}$Te (barrier) are $0.6462\nm$ and $0.6474\nm$, respectively. This leads to a strain tensor $\epsilon=\diag(\epsilon_\parallel,\epsilon_\parallel,\epsilon_\perp)$ with $\epsilon_\parallel$ and $\epsilon_\perp$ given in Table~\ref{table_kdotpcoeff}.


\bibliographystyle{apsrev4-2}
\bibliography{_HgTe_symm}


\end{document}